# Effect of the nanowire diameter on the linearity of the response of GaN-based heterostructured nanowire photodetectors

Maria Spies[1], Jakub Polaczyński[1,*], Akhil Ajay[2], Dipankar Kalita[2], Jonas Lähnemann[2,†], Bruno Gayral[2], Martien I. den Hertog[1], Eva Monroy[2]

[1] University Grenoble-Alpes, CNRS, Institut Néel, 25 av. des Martyrs, 38000 Grenoble, France

[2] University Grenoble-Alpes, CEA, INAC, 17 av. des Martyrs, 38000 Grenoble, France

OrcIDs:
Martien I. den Hertog: 0000-0003-0781-9249
Eva Monroy: 0000-0001-5481-3267
Maria Spies: 0000-0002-3570-3422
Jonas Lähnemann: 0000-0003-4072-2369
Akhil Ajay: 0000-0001-5738-5093

## Abstract

Nanowire photodetectors are investigated because of their compatibility with flexible electronics, or for the implementation of on-chip optical interconnects. Such devices are characterized by ultrahigh photocurrent gain, but their photoresponse scales sublinearly with the optical power. Here, we present a study of single-nanowire photodetectors displaying a linear response to ultraviolet illumination. Their structure consists of a GaN nanowire incorporating an AlN/GaN/AlN heterostructure, which generates an internal electric field. The activity of the heterostructure is confirmed by the rectifying behavior of the current-voltage characteristics in the dark, as well as by the asymmetry of the photoresponse in magnitude and linearity. Under reverse bias (negative bias on the GaN cap segment), the detectors behave linearly with the impinging optical power when the nanowire diameter is below a certain threshold (≈ 80 nm), which corresponds to the total depletion of the nanowire stem due to the Fermi level pinning at the sidewalls. In the case of nanowires that are only partially depleted, their nonlinearity is explained by a nonlinear variation of the diameter of their central conducting channel under illumination.

[*] Current address: Institute of Physics Polish Academy of Sciences, Aleja Lotnikow 32/46, 02668 Warsaw, Poland, and International Research Centre MagTop, Aleja Lotnikow 32/46, 02668 Warsaw, Poland.
[†] Current address: Paul-Drude-Institut für Festkörperelektronik, Leibniz-Institut im Forschungsverbund Berlin e.V., Hausvogteiplatz 5-7, 10117 Berlin, Germany

# 1. Introduction

Nanowire photodetectors [1–5] have the potential to surpass the spatial resolution and speed of planar devices, as well as having intrinsic advantages such as the reduced dimensions and small electrical cross section. In the ultraviolet region, ZnO and GaN nanowires have been intensively studied as spectrally-selective photodetectors. For this application, III-nitride nanowires present advantages in terms of robustness and heterostructuring possibilities. Nanowire photoconductors are characterized by high photocurrent gains, which can reach $10^6$, and strong spectral contrast above and below the bandgap. A general feature in nanowire photoconductors is the fact that the photocurrent scales sublinearly with the impinging laser power, which has been shown for single GaN nanowires regardless of the presence of heterostructures [6–10], as well as for nanowires of other material systems such as ZnTe [11], ZnO [12,13], InP [14], CuO [15], and GaAs [16]. This sublinearity of the response hampers the use of such devices for quantification of the radiant fluence, and restricts their application domain to digital detection. The high photocurrent gain and the sublinearity have been related to the light-induced reduction of the depletion layer at the nanowire sidewalls [3,6,7,12,17]. Indeed, the large surface to volume ratio in nanowires makes them very sensitive to surface effects (presence of charge traps or Fermi level pinning, which can be modified by adsorbed species). In the case of undoped GaN nanowires, Sanford et al. reported an improvement of the linearity in nanowires with small diameter ($\approx$ 100 nm), which they attributed to the total depletion of the nanowires associated with the axial electric field generated by asymmetric Schottky-like contacts [6].

In this paper, we systematically investigate GaN nanowire photodetectors with an embedded AlN/GaN/AlN heterostructure, which is responsible for creating an axial electric field within the nanowire. We demonstrate that for reverse-biased (negative bias on the GaN cap segment) nanowires with diameters < 80 nm, the response to ultraviolet illumination is in fact linear, which is consistent with the total depletion of the nanowire stem due to the Fermi level pinning at the sidewalls. Using theoretical calculations, we show that in these thin wires illumination does not have any significant effect on the lateral electric field. On the contrary, in large nanowires (diameter > 80 nm), which are only partially depleted, illumination results in a variation of the lateral electric field and a nonlinear increase of the conducting section in the center of the wires. Therefore, we show here that the depletion of the nanowire due to surface effects can be beneficial for the photodetector performance since it allows the fabrication of linear devices if the nanowire heterostructure is properly designed.

## 2. Sample Design, Growth and Device Fabrication

The samples under study are GaN nanowires incorporating an AlN/GaN/AlN heterostructure to enhance the responsivity at low bias, as a consequence to the internal electric field generated by the large difference in spontaneous and piezoelectric polarization between GaN and AlN. A schematic description of the structure is presented in figure 1(a). For the design of the heterostructure, we calculated the band profile using the Nextnano[3] 8×8 **k·p** self-consistent Schrödinger-Poisson solver [18] using the parameters listed in ref. [19]. The result is illustrated in figure 1(c). The heterostructure consists of two 10-nm-thick AlN insertions, separated by 2.3 nm of GaN. The heterostructure is surrounded by segments of undoped GaN (each 130 nm long), while the ends of the nanowires are doped at $8\times10^{17}$ cm$^{-3}$ to facilitate Ohmic contacts. Thereby, the depletion region induced by the internal electric field is maximized. As self-assembled nanowires grow along the [000−1] crystallographic axis [20], the polarization-induced depletion region is located below the heterostructure. In turn, accumulation of free electrons occurs on top of the heterostructure. To avoid the risk of covering the depletion region when depositing the contacts, the wire is asymmetric, with the heterostructure located towards the top of the nanowire. Under illumination, the depletion region is expected to separate charge carriers, as illustrated in figure 1(c).

The presence of the GaN/AlN heterojunction favors the collection of photogenerated electrons, but it is an obstacle for hole transport. The field-emission transport through such a barrier should be negligible. However, transport through relatively large AlN barriers has been experimentally observed [21,22]. The strong band bending in the heterostructure favors a generation-recombination current that involves holes from the stem and electrons from the cap region. In the case of a single barrier, transport proceeds via interband Zener tunneling [21]. In our case, where two barriers are involved, the states in the intermediate GaN section should play a role as intermediate states in the transport process.

Note that the thickness of the layers in the AlN/GaN/AlN heterostructure is not critical, within certain limits. Increasing the size of the AlN barriers would have the positive impact of decreasing the dark current, but it would lead to problems for the collection of photogenerated holes. Therefore, if a larger heterostructure were desired, it should be implemented as an AlN/GaN multi-quantum-well structure. However, if several periods were added, the heterostructure would start showing its own contribution to the spectral response, particularly when positive bias is applied to the cap. On the other hand, a thinner heterostructure would result in a reduction of the electric field in the stem. This would decrease the extension of the depletion region where the charge carrier separation takes place.

These GaN nanowires were synthesized by plasma-assisted molecular beam epitaxy (PAMBE) on Si(111) substrates. The growth rate of the GaN nanowires was ≈ 0.11 nm/s and the substrate temperature was $T_S = 810°C$. Prior to the growth of the nanowires, an AlN buffer layer was deposited using the 2-step growth procedure described in ref. [23]. In the sample under study, the nanowire base consists of a 2.5-μm-long GaN stem doped with Ge ([Ge] = $8 \times 10^{17}$ cm$^{-3}$), and a 130-nm-long undoped GaN segment. This was followed by the AlN/GaN/AlN (10 nm/2.3 nm/10 nm) heterostructure and by a 130-nm-thick undoped GaN segment, which is finally capped with a Ge-doped ([Ge] = $8\times10^{17}$ cm$^{-3}$) GaN segment of about 410 nm length. The choice of Ge as n-type dopant was motivated by previous reports indicating that Si has a tendency to migrate to the nanowire surface [24], whereas Ge incorporates efficiently without deformation of the nanowire geometry [25,26]. During the growth of the AlN barriers, a thin AlN shell forms around the GaN stem and the AlN/GaN/AlN heterostructure, as indicated in the scheme in figure 1(a). A scanning electron microscopy (SEM) image of the as-grown nanowire ensemble can be seen in figure 1(b), where the arrows on the side of the SEM image mark the location of the heterostructure. The nanowires exhibit diameters of ≈ 80 nm.

The as-grown nanowire ensemble is sonicated in isopropanol and dispersed on sets of Si$_3$N$_4$ membranes fabricated on Si(100). The aim of the membranes is to provide an electron-transparent support to enable scanning transmission electron microscopy (STEM) studies of the nanowires after their optoelectronic characterization. Such SiN$_x$ grids are fabricated using a 400-μm-thick n$^{++}$ Si(100) wafer with layers of 200 nm SiO$_2$ and 40 nm of stoichiometric Si$_3$N$_4$ on both sides. Using laser lithography, membrane windows are defined on one side of the wafer and subsequently etched through the nitride and oxide layers using reactive ion etching. Then, a KOH bath etches through the Si and the oxide on the other side of the wafer, leaving only the 40-nm-thick layer of SiN, which then constitutes the 200 μm × 200 μm membrane. In a further step, contact pads and markers are defined by laser lithography and electron beam deposition of Ti/Au (10 nm/35 nm). The nanowires dispersed on such membranes are either bundles [figure 1(d)] or single wires [figure 1(e)], which are contacted using electron beam lithography and deposition of Ti/Al (10 nm/120 nm). The example of a contacted nanowire can be seen in figure 1(f).

## 3. Experimental results

The current-voltage (I-V) characteristics in the dark exhibit a strongly rectifying behavior, as illustrated in figure 2 for two typical specimens. The asymmetry of the I-V curve is explained by the electronic asymmetry presented in figure 1(c), which results from the polarization fields in wurtzite III-nitride heterostructures. The band profile resembles that of a Schottky diode, where the cap layer plays the role of the metal Schottky contact. Consistently, forward bias in figure 2 corresponds to a positive voltage being applied to the nanowire cap with respect to the stem.

Following the interpretation for the case of a single AlN insertion [20], the reverse current is associated to a leakage path through the barriers, involving a GaN shell, surface conduction or the coalescence of multiple wires [10]. Under forward bias, among the set of nanowires under study, we observe a strong dispersion of the dark current, varying from a few nanoamperes to microamperes at +1 V bias. STEM images of 10 nanowire bundles (19 nanowires) show no correlation of the dark current with the number of nanowires that are effectively contacted or with possible fluctuations in the thickness of the AlN barriers. However, there is a correlation between the dark current and the nanowire diameter, as illustrated in the inset of figure 2. For clarity, we have divided the studied nanowires into two groups: those with dark current in the nanoampere range at +1 V bias – we shall call them group A from here on and those that display microampere-ranged currents at the same bias – we shall call them group B. As shown in the inset of figure 2, the limit between the two groups is found for a diameter of ≈ 80 nm.

Moving to measurements under illumination, the variation of the photocurrent as a function of the ultraviolet irradiance was studied at 325 nm. Figures 3(a) and (b) present typical results for specimens in groups A and B, respectively. They were measured at zero and negative bias. Straight lines are fits to $I_{ph} = AP_{opt}^{\beta}$, where $I_{ph}$ is the photocurrent, $P_{opt}$ is the impinging optical power and the proportionality constant $A$ and the power law exponent $\beta$ are fitting parameters. Note that $\beta = 1$ indicates that the photoresponse is linear. In both figures, the photocurrent at zero bias scales sublinearly with the impinging irradiance, which is consistent with previous reports on samples containing GaN/AlN superlattices [27]. This behavior differs from the observations in planar photodetectors [28]. Planar photovoltaic devices are systematically linear since the photocurrent is due to the linear generation of electron-hole pairs separated by the internal electric field. The deviation from this behavior in the case of nanowires reveals the involvement of an additional mechanism in the zero-bias photoresponse, which remains unknown at this moment.

Under −100 mV bias, nanowires in groups A and B exhibit different behaviors. Figure 3(a) displays a linear photocurrent behavior for group A, whereas the photoresponse of group B in figure 3(b) remains sublinear. Additionally, figure 3(a) shows a significant improvement of the linearity for bias voltages as low as −10 mV. Figure 3(c) displays a summary of the values of $\beta$ (at −100 mV) as a function of the dark current through the nanowire (at +1 V) for all the nanowires in this study. The nanowires in group A present $\beta = 1.0 \pm 0.2$, i.e. their photocurrent scales linearly with the optical power within the error bars of the measurement, whereas the nanowires in group B clearly show a sublinear behavior, with $\beta = 0.45 \pm 0.11$.

Note that, for this linearity measurements, the range of bias voltage applied to the nanowires was chosen to keep the maximum photocurrent lower than 10 μA, to prevent device failure. This implies that the maximum applied voltage was in the range of −1 V (lower for some of the specimens). In the devices where a linear behavior is observed (group A), the linearity improves with bias, without any indication of degradation at higher bias.

To verify that the photocurrent induced by ultraviolet illumination stems from the GaN nanowires, we have recorded the spectral response for both sets of devices. The results are presented in figure 3(d). In both cases (groups A and B), the absorption exhibits a sharp cutoff around 365 nm, which corresponds well to the band gap of GaN at room temperature. This experiment confirms that in both cases a potential leakage photocurrent through the silicon substrate is negligible.

If we approximate the exposed photodetector area by the in-plane cross-section of the contacted nanowire (on average, 1.5 µm × nm), we can estimate that the typical responsivity (geometric mean) for an irradiance of 10 mW/cm is around 0.3 A/W at zero bias. (Regarding the calculation of the responsivity, see section 1 of the Supporting Information). At a bias of −100 mV, the typical responsivity, measured under the same conditions and calculated in the same manner, increases to around 20 A/W for nanowires in group A, and up to around 700 A/W for nanowires in group B. The increase of the responsivity with the nanowire diameter is consistent with previous reports on GaN nanowire photodetectors [6,30].

To confirm the role of the heterostructure in the responsivity of the nanowires, we have compared the photocurrent under forward and reverse bias. Under reverse bias, the response is expected to be dominated by the presence of the space charge region, which separates photogenerated electrons and holes. The response is hence expected to resemble that of a Schottky diode (low dark current and linear response with the optical power) [31]. In contrast, under forward bias, the space charge region disappears and the nanowire resembles a photoresistor (high dark current and sublinear response) [31]. Figure 4 presents the variation of the photocurrent as a function of the ultraviolet irradiance in a specimen from group A measured at +1 V and −1 V bias. As expected, the photoresponse scales linearly with the irradiance under reverse bias ($\beta = 0.96 \pm 0.06$) only, whereas forward bias results in a strongly sublinear behavior ($\beta = 0.61 \pm 0.05$). This asymmetric behavior is a confirmation of the role of the AlN/GaN/AlN heterostructure in the photoresponse.

Finally, we have assessed the effect of the limited response time of the photodetectors on our measurements. The measurements reported above were recorded using a synchronous detection setup (see Methods for details) where the light is chopped at a frequency of 33 Hz. In the case of planar structures, it is known that the chopping frequency has dramatic effects on the linearity and spectral response of photoconductors [28,31], whereas Schottky photodiodes are relatively insensitive to the chopping frequency in the typical experimental range (1-1000 Hz). However, in the case of single GaN nanowires, we have reported that the spectral response does not vary as a function of the chopping frequency [7]. To validate the results of this manuscript, we have verified that the value of the β exponent as a function of the irradiance is insensitive to the chopping frequency. The experimental confirmation is presented as section 2 of the Supporting Information.

## 4. Discussion

The drastic reduction of the dark current in nanowires with a diameter below ≈ 80 nm has been observed previously in GaN nanowires [30], and it was explained by the presence of a space charge layer extending inwards from the nanowire sidewalls. In the report by Calarco et al. [30], total depletion of the GaN nanowires was obtained for a diameter of 85 nm, when the residual doping level was $6.25 \times 10^{17}$ cm$^{-3}$. To confirm that our result is consistent for the doping level in the nanowires under study, three-dimensional calculations of the band diagram have been carried out. Different diameters, namely 50, 60, 80 and 120 nm, were considered. The results of the simulations are summarized in figure 5(a), which displays the cross-sectional view of the conduction band structure in the doped stem region extracted 200 nm below the first GaN/AlN heterointerface [see dashed line in figure 1(a)]. We note that for nanowires with a diameter of 60 nm, the space charge regions extending from opposite sidewalls touch each other, and the location of the conduction band edge in the center of the nanowire increases by about 100 meV when decreasing the nanowire diameter from 80 to 50 nm. This confirms the full depletion of the thin nanowires and justifies the drastic drop in the dark current.

Ultraviolet illumination is known to unpin the Fermi level at the nanowire sidewalls. This phenomenon has been experimentally studied by Pfüller et al. [29], and it was attributed to photoinduced desorption of oxygen from the nanowire sidewalls. Therefore, to simulate the effect of ultraviolet illumination, we have analyzed the consequences of changing the position of the Fermi level at the surface in the range of 2.0 eV to 2.2 eV below the conduction band edge of the AlN shell. Taking a look at the simulations of a nanowire with a diameter of 50 nm (group A) [figure 5(b)] we observe that changes to the Fermi level pinning shift the radial position of the conduction band as a whole across the nanowire, but the shape of the potential profile is not modified. In other words, the component of the electric field along the nanowire diameter seen by photogenerated electrons is approximately the same in all cases, with its maximum value at the GaN/ AlN interface being 210 kV/cm ± 3%. In a nanowire with diameter of 120 nm (group B) [figure 5(c)] we note a different behavior. When the location of the Fermi level pinning changes from 2.2 eV to 2.0 eV below the conduction band edge of the AlN shell, the maximum radial electric field varies from 360 kV/cm to 240 kV/cm (by more than 30%). At the same time, the space charge region at the sidewalls of the nanowire shrinks, increasing the extent of the central conducting channel in the nanowire. Therefore, in thick nanowires (group B), light induces not only a linear increase in the carrier concentration, but also a nonlinear variation in the diameter of the conducting channel that such carriers have to traverse to be collected.

This explains also the enhancement of the responsivity with the nanowire diameter. The responsivity is linked to the total number of photogenerated carriers, i.e. it should increase with the square of the nanowire radius. To this dependence, we have to add the variation of the conductivity due to the change in the diameter of the central conducting channel in the stem. Both phenomena are relatively independent. In small, fully depleted nanowires, the variation of the responsivity with the diameter will be given by the change in the total amount of photogenerated carriers. In large, partially-depleted wires, it is the modulation of the conductive section that dominates, which can lead to huge photocurrent gains. A theoretical analysis of both contributions can be found in ref. [32].

In summary, these different behaviors of the radial potential profiles explain the observed differences in linearity ($\beta$) as a function of the nanowire diameter.

In this manuscript, linearity is observed for nanowires with a diameter smaller than $\approx$ 80 nm. For larger nanowires, there are two approaches to improve the linearity, namely obtaining a full depletion of the nanowires or rendering the band bending at the sidewalls insensitive to light. Full depletion of larger nanowires can be achieved by reducing the doping level. Reducing the sensitivity of the band bending to light is more challenging. The use of a thicker AlN shell might help, but there is a risk of generating structural defects due to the lattice mismatch between m-plane AlN and GaN. Alternatively, the use of dielectrics as passivation layer should be explored.

## 5. Conclusions

We have demonstrated single-nanowire ultraviolet photodetectors consisting of a GaN nanowire with an embedded AlN/GaN/AlN heterostructure, which generates an electric field along the nanowire axis as a result of the difference in polarization between III-nitride compounds with wurtzite crystal structure. The influence of the heterostructure is confirmed by the rectifying behavior of the current-voltage characteristics in the dark, and by the asymmetry of the photoresponse in magnitude and linearity. Under reverse bias (negative bias on the cap segment), the detectors behave linearly with the impinging optical power when the nanowire diameter remains below a certain threshold ($\approx$ 80 nm). This is explained by the linearity of the photogeneration process, the separation of photogenerated carriers induced by the axial electric field, and the fact that illumination does not have a significant effect on the radial electric field. In the case of nanowires that are not fully depleted (diameter > 80 nm), the light-induced change in the Fermi level at the sidewalls results in a variation of the diameter of the central conducting channel in the stem, which leads to an overall nonlinear photoresponse.

## 6. Methods

The structural properties of the nanowires were probed both by high angle annular dark field (HAADF) scanning transmission electron microscopy (STEM) and TEM using a probe corrected FEI TITAN Themis working at 200 kV and a CM 300 working at 300 kV. We used a DENSSolutions 6 contact double tilt holder. Current voltage (I-V) characteristics have been investigated with the nanowires connected to an Agilent 4155C semiconductor parameter analyzer. Forward bias corresponds to higher voltage in the cap segment than in the stem. Photocurrent measurements as a function of optical power have been carried out using an unfocused continuous-wave HeCd laser (wavelength = 325 nm, spot diameter on the sample ≈ 1 mm), chopped at 33 Hz (unless indicated). The nanowire is biased and connected in series with a ×$10^6$ V/A transimpedance amplifier, which is read out by a Stanford Research Systems SR830 lock-in amplifier. The spectral response has been measured using the same read-out configuration, but exciting with light from a 450 W xenon lamp passed through a Gemini 180 Jobin-Yvon monochromator. In general, the bias was chosen to keep the maximum photocurrent lower than 10 µA to prevent device failure. All measurements were carried out at room temperature.

Theoretical calculations of the band diagram were performed using the commercial 8×8 **k·p** band Schrödinger-Poisson equation solver nextnano[3]. The nanowire was modeled as a hexahedral prism consisting of a 150 nm long n-type GaN segment followed by 130 nm of undoped GaN, an AlN/GaN/AlN (10 nm/2.3 nm/10 nm) heterostructure, 130 nm of undoped GaN, and 50 nm of n-type GaN. The n-type doping density and residual doping were fixed to $8\times10^{17}$ cm$^{-3}$ and $1\times10^{17}$ cm$^{-3}$, respectively. The structure was defined on a GaN substrate to provide a reference in-plane lattice parameter, and was embedded in a rectangular prism of air to include elastic strain relaxation. In a first stage, the three-dimensional strain distribution was calculated by minimization of the elastic energy assuming zero stress at the nanowire surface. Then, for the calculation of the band profiles, the piezoelectric fields resulting from the strain distribution were taken into account. Diameters of 50, 60, 80 and 120 nm were considered. Regarding the treatment of the surface, it is generally accepted that in GaN nanowires the Fermi level at the m-plane sidewalls is located around 0.6 eV below the conduction band edge [33,34] with a certain dependence on the environment [17,35]. However, the presence of an AlN/GaN/AlN heterostructure leads to the formation of an AlN shell around the GaN stem, which places the Fermi level around 2.1 eV below the conduction band of AlN [36]. Therefore, in our calculations, we have fixed the Fermi level at the AlN sidewalls of the stem at 2.1 eV below the AlN conduction band. On the contrary, in the cap region, we have fixed the Fermi level at the GaN/air interface at 0.6 eV below the conduction band. However, this latter value has no critical influence on the results, since the area of the cap exposed to light is small and the polarization-induced accumulation of electrons at the upper AlN/GaN heterointerface screens the effect of the surface.

## Acknowledgements

Financial support from the ANR-COSMOS (ANR-12-JS10-0002) project and the AGIR 2016 Pole PEM funding proposed by Grenoble Alpes University (UGA) for the CoPToN project is acknowledged. J.P. acknowledges support of the Erasmus+ programme of the European Union. A.A. acknowledges financial support from the French National Research Agency via the GaNEX program (ANR-11-LABX-0014). We benefitted from the access to the Nano characterization platform (PFNC) in CEA Minatec Grenoble. Membrane production and nanowire contacting have been carried out at the NanoFab cleanroom of Institut Néel, Grenoble. Thanks are due to Bruno Fernandez and Jean-François Motte for cleanroom support, as well as to Yohan Curé and Yann Genuist for PAMBE support.

**Figure captions**

**Figure 1.** (a) Schematic of the nanowires under study. (b) SEM image of the as-grown nanowire ensemble. The position of the insertion can be identified on the top quarter of the nanowires as the growth of the strained GaN on the AlN insertion results in a slight reduction of the nanowire diameter. (c) One-dimensional simulation of the photoactive part of the nanowire showing the band bending due to the AlN insertions. Electron-hole pairs photogenerated in the space charge region will get separated by the internal electric field, as depicted schematically. (d-e) HAADF STEM images of a bundle of nanowires and a single nanowire, respectively. (f) SEM image of a contacted nanowire.

**Figure 2.** Current-voltage characteristics of two typical nanowires, one with small diameter <80 nm (green) and one with diameter >80 nm (orange). Inset: Dark current at +1 V bias as a function of the diameter of the nanowires measured by STEM in proximity of the AlN/GaN/AlN insertion. The error bars account for the different diameters of nanowires in a bundle. The dotted line is a guide to the eye. Nanowires with diameters <80 nm respond in the nA range (group A), whereas nanowires with diameters >80 nm present a dark current in the µA range (group B).

**Figure 3.** Photocurrent measurements as a function of the irradiance (impinging laser power per unit of surface) at 325 nm for (a) a typical group-A nanowire (diameter < 80 nm), and (b) a typical group-B nanowire (diameter > 80 nm). Bias is indicated in the legends. Lines are fits to $I_{ph} = AP_{opt}^{\beta}$, where $I_{ph}$ is the photocurrent, $P_{opt}$ is the impinging optical power and $A$ and $\beta$ are fitting parameters. The values of $\beta$ are indicated in the figure. (c) Variation of $\beta$ as a function of the dark current (measured at +1 V). Note, the correlation of almost-linear nanowires with dark current in the nA range, and clearly sublinear nanowires with dark current

in the µA range. The solid line is a guide to the eye. (d) Spectral response measurements for typical group-A and group-B nanowire specimens. The dashed line marks the wavelength of the GaN band gap at room temperature.

**Figure 4.** Photocurrent measurements as a function of the irradiance at 325 nm for a typical group-A nanowire measured under forward and reverse bias.

**Figure 5.** Cross-sectional view of the band structure in the doped stem region of the nanowire, 200 nm below the undoped region (a) for nanowire diameters of 50, 60, 80, 120 nm, with the Fermi level pinned 2.1 eV below the conduction band edge of AlN. With decreasing diameter, the difference in energy between Fermi level and the lowest point in the conduction band increases from 38 meV to 140 meV. In (b) and (c), the Fermi level pinning is varied between 2.0, 2.1 and 2.2 eV for nanowires with diameters of 50 nm and 120 nm, respectively.

**Figure 1**

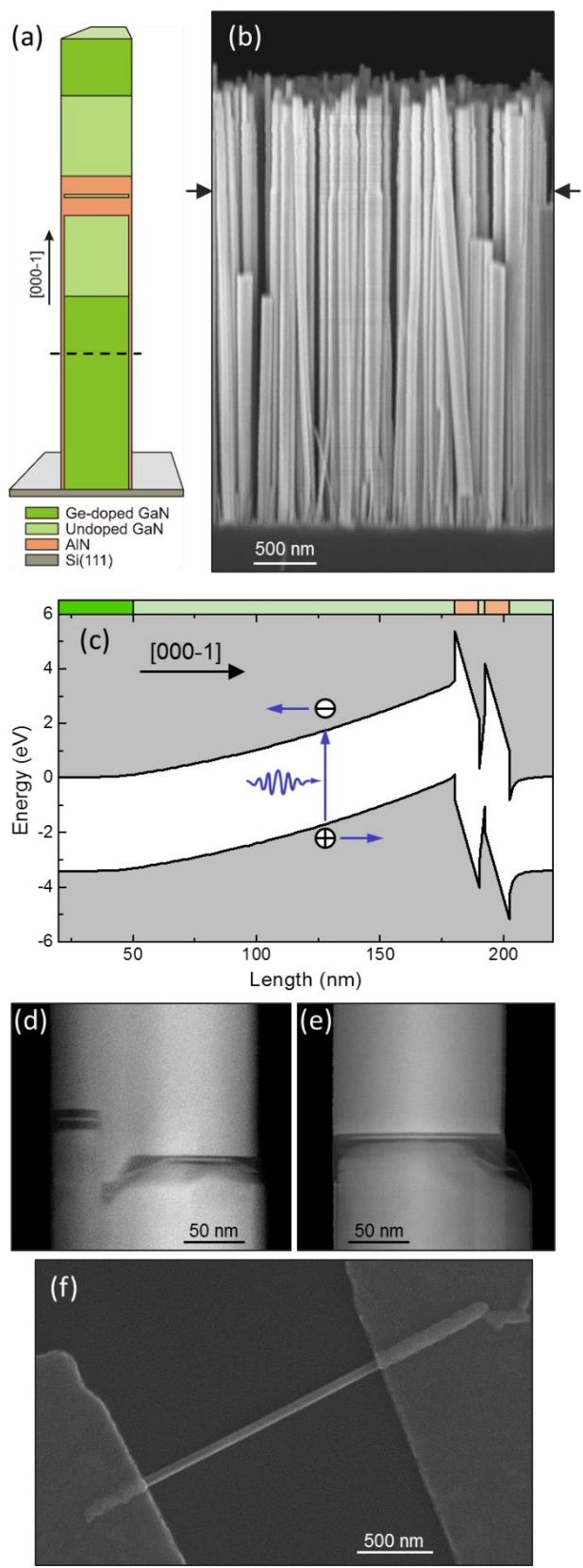

**Figure 2**

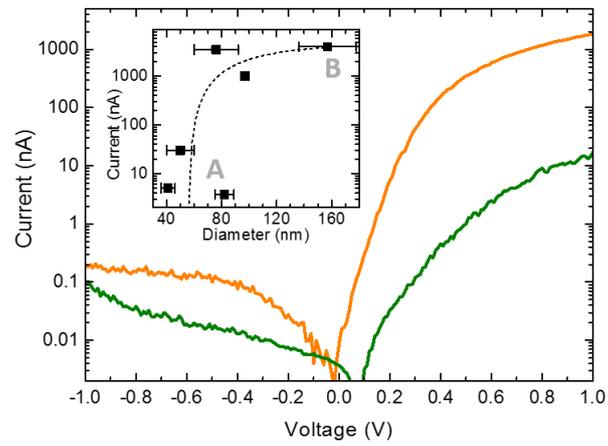

**Figure 3**

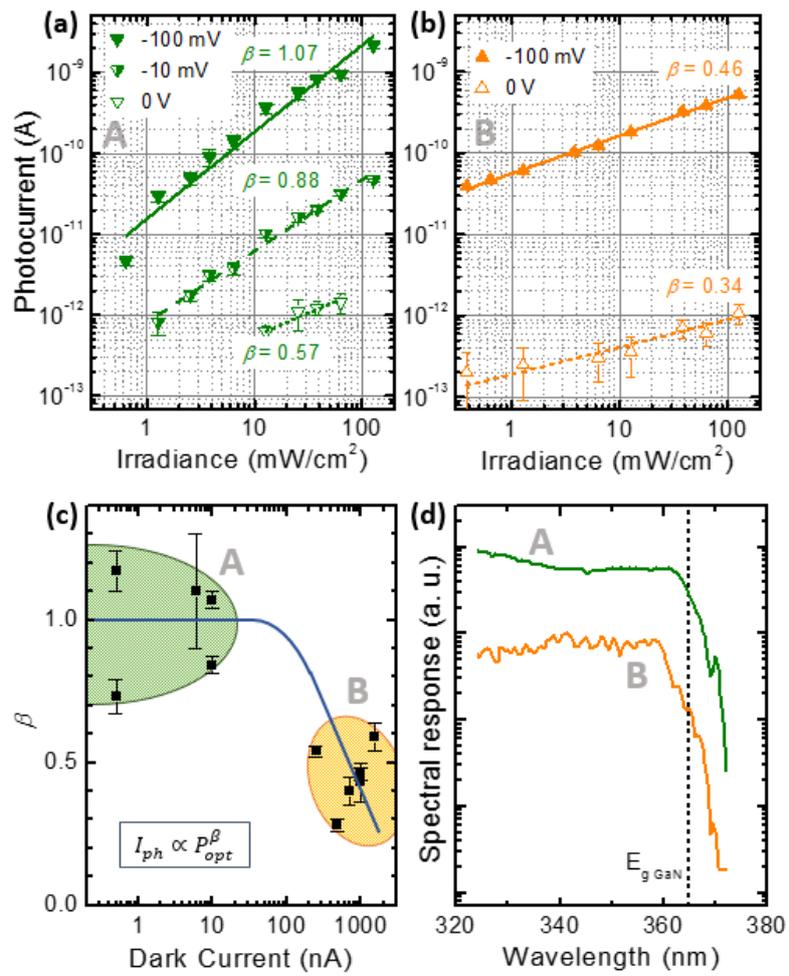

**Figure 4**

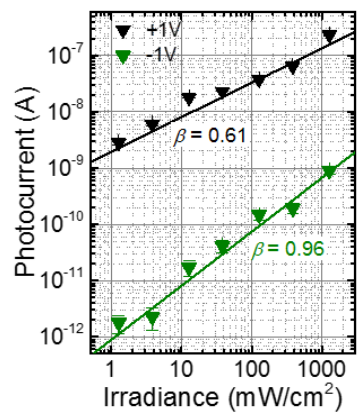

**Figure 5**

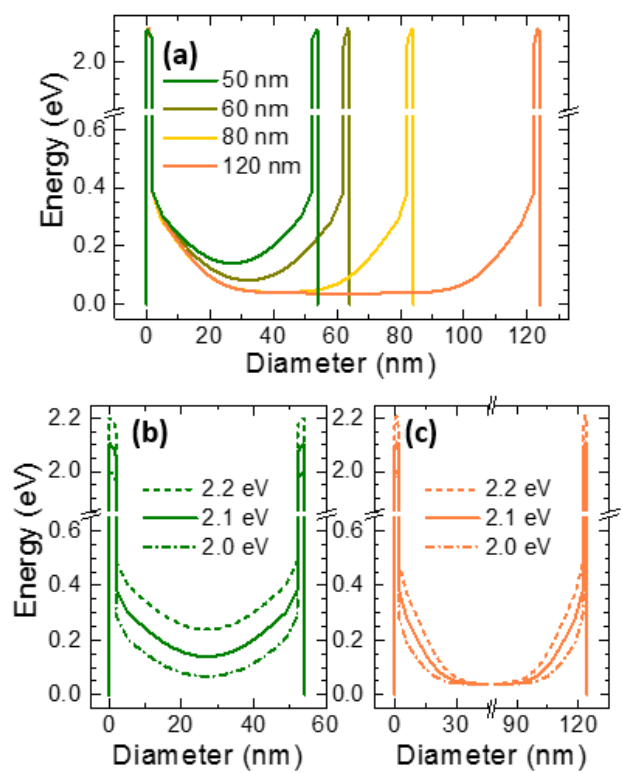